\documentstyle[openbib,12pt,fleqn]{article}
\renewcommand{\baselinestretch}{1.2}

\newcommand{\fg}[1]{\item {\label{#1}}}

\newcommand{\rf}[1]{\raisebox{1ex}{\cite{#1}}}

\newcommand{\barr}{\begin{array}}
\newcommand{\bea}{\begin{eqnarray}}
\newcommand{\beq}{\begin{equation}}
\newcommand{\ear}{\end{array}}
\newcommand{\eea}{\end{eqnarray}}

\newcommand{\eeq}{\end{equation}}

\newcommand{\spao}[1]{\mbox{\hspace{#1}}}
\newcommand{\spav}[1]{\parbox{1mm}{\vspace*{#1}}}

\newcommand{\ssty}{\scriptstyle}
\newcommand{\sssty}{\scriptscriptstyle}

\newsavebox{\ipiu}
\newsavebox{\imen}

\sbox{\ipiu}{$\ssty i \sssty +1$}
\sbox{\imen}{$\ssty i \sssty -1$}

\begin{document}
\bibliographystyle{unsrt}

\begin{titlepage}
\centering \spav{1cm}\\
{\LARGE\bf Periodic Orbit Theory of\\ }
{\LARGE\bf Anomalous Diffusion\\ }
\spav{2cm}\\
{\large Roberto Artuso\\}
{\normalsize\em Dipartimento di Fisica dell'Universit\'a and I.N.F.N.\\
Via Celoria 16, I--20133 Milano, Italy\\} 
{\large Giulio Casati\\}
{\normalsize \em Dipartimento di Fisica dell'Universit\'a and I.N.F.N.\\
Via Castelnuovo, I--22100 Como, Italy\\} 
{\large and\\} 
{\large Roberto Lombardi\\}
{\normalsize \em Dipartimento di Fisica dell'Universit\'a\\
Via Celoria 16, I--20133 Milano, Italy\\}
\spav{1mm}\\
\vfill
{\centerline \small PACS numbers: 05.45.+b,05.60.+w,05.40.+j}
\vfill
{\small\bf Abstract\\}
\spav{2mm}\\
{\small\parbox{13cm}{\spao{4mm}
We introduce a novel technique to find the asymptotic time behaviour 
of deterministic systems exhibiting anomalous diffusion. The procedure 
is tested for various classes of simple but physically relevant 
1--D maps and possible relevance of our findings for 
more complicated problems is briefly discussed.
}}
\spav{4mm}\\
\end{titlepage}
\setcounter{footnote}{0}
In the last few years the phenomenon of deterministic diffusion has 
been widely investigated: as a matter of fact a major theoretical 
challenge in the study of dynamical systems is to understand 
thoroughly the generation of typical stochastic properties in purely 
deterministic systems. Moreover even the simplest maps for which 
deterministic diffusion has been observed are supposed to model
the behaviour of relevant 
physical systems (like Josephson junctions in presence of a microwave 
field, see e.g. {\cite{J}}).
A recent approach, introduced independently in 
{\cite{DA}} and {\cite{DCEG}},
leads to an expression for 
the diffusion coefficient in terms of the periodic orbits of the 
system, in the form of a cycle 
expansion{\rf{AAC1}}. 
Cycle expansions have been applied in a number 
of different contexts (see for instance {\cite{AAC2,CE}}): they work 
remarkably well for low dimensional hyperbolic systems, 
provided their topology (symbolic dynamics) is under control.
The problem of controlling the topology of the system is well 
illustrated, in the context of diffusion properties via cycle 
expansions,
when one deals with the 
infinite Lorentz gas with bounded horizon (see {\cite{LGC}}). 
If 
one relaxes the hypothesis of pure hyperbolicity (absence of marginal 
stability),
much more care has to be taken, and generally the 
effectiveness of the expansions is regained only if one is able to sum 
infinite contributions shadowing the marginal fixed point{\rf{AAC2}}. 
We emphasize that this problem is of paramount relevance, as it 
naturally arises when dealing with generic hamiltonian systems, in 
which elliptic islands of stability and hyperbolic homoclinic webs 
coexist.

Here we address explicitly the problem of marginal stability, which is 
tightly connected to the appearance of anomalous diffusion (and also 
represents a crucial feature in establishing a good theory for 
diffusion in generic two--dimensional area preserving 
maps{\rf{BSM}}). 
We extend the theory developped in {\cite{DA,DCEG}} to study 
deviations from normal diffusion, expressing the asymptotic time 
behaviour in terms of properties of the periodic orbits of the system. 
The method is then applied to a class of 1--D maps for which anomalous 
diffusion has been previously observed.

We begin by recalling the essential features of the periodic orbit 
theory of normal diffusion{\rf{DA,DCEG}}, by considering the simplest 
context in which it may be applied.
We will consider lifts of one--dimensional circle maps
\beq
x_{t+1}=f(x_t)\,\,t\in {\bf N}\quad f(x+n)=n+f(x) \quad f(-x)=-f(x)
\label{tmap}
\eeq
together with corresponding torus maps $x_{t+1}=f(x_t) 
|_{mod\,1}=\hat{f}(x_t)$. Normal diffusion means that asymptotically 
$\sigma^2 (t)=<(x_t-x_0)^2>\sim 2Dt$, where the average is over 
initial conditions (by symmetry $x_0$ may be taken in the unit 
interval). The set of all periodic orbits of $\hat{f}$ will be denoted 
by $\{p\}$: each orbit will be characterized by its period $n_p$
stability $\Lambda _p$ (product of the derivatives along the cycle) and 
integer winding number $\sigma_p$ (such that for each cycle point $x_{i(p)}$ 
we have $f^{n_p} (x_{i(p)})=x_{i(p)}+\sigma_p$).
We focus our attention on the generating function 
$<e^{\beta(x_t-x_0)}>=\Omega_t(\beta)$; it has been shown{\rf{DA,DCEG}} that 
its asymptotic behaviour is 
dominated by the leading eigenvalue of an appropriate transfer 
operator: $\Omega_t(\beta)\sim z(\beta)^{-t}$, where 
$z(\beta)$ is the smallest solution of
\[
\zeta_0^{-1}(z(\beta),\beta)\,=\,\prod_{\{p \} }\left( 1- 
\frac{z^{n_p}\exp(\sigma_p \beta)}{|\Lambda_p|}\right)\,=\,0
\]
Expansion of the generating function for small $\beta$ thus yields
\[
D=-\frac{1}{2}\left. \frac{d^2}{d\beta^2} z(\beta) \right|_{\beta=0}
\]

Anomalous diffusion is associated to a vanishing or diverging $D$. 
In order to generalize the theory we take into account that, up to time $t$, 
$\Omega_t(\beta)$ gets contributions from periodic orbits with 
$n_p\leq t${\rf{DA,DCEG}} (as it is connected 
to the trace of the t--th power of the 
transfer operator): we thus may write in general
\beq
\Omega_t(\beta)\sim \left[ z^{[t]}(\beta) \right]^{-t}
\label{conj}
\eeq
where 
$z^{[t]}(\beta)$ is such that 
$\zeta_{0[t]}^{-1}(z^{[t]}(\beta),\beta)=0$, $\zeta_{0[t]}^{-1}$ 
denoting the cycle expansion of $\zeta_0^{-1}$ truncated to finite order $t$.
From (\ref{conj}) we thus get the asymptotic behaviour
\beq
\sigma^2(t)\,\sim \, -t \left.\frac{\partial^2 /\partial \beta^2\, 
\zeta_{0[t]}^{-1} (z,\beta)}{\partial / \partial z \, 
\zeta_{0[t]}^{-1} (z,\beta)}\right|_{z=1,\beta=0}
\label{ast}
\eeq
where we took into account probability 
conservation{\rf{AAC1}} ($z^{[t]}(0) \to 1$ as $t \to \infty$) as well as 
symmetry of the map.
This procedure may be alternatively interpreted as follows: for a 
sequence of large increasing times $t_k$ we approximate the system by 
a sequence of hyperbolic (regularly diffusing) systems $S_k$ with 
polynomial zeta functions 
$\zeta_0^{-1}(z,\beta)_{S_k}=\zeta^{-1}_{0[t_k]}(z,\beta)$: 
$\sigma^2(t)\sim 2D(t_k)\cdot t$ for $t\stackrel{<}{\sim} t_k$ (for both the 
original system and $S_k$), thus corrections to normal diffusion are 
connected with the asymptotic behaviour of $D(t)$.

To test the theory we consider a map of the form (\ref{tmap}) (see 
fig. 1) characterized by the presence of a marginal
fixed point{\rf{GT,GNZ}}.
The corresponding map on the torus $\hat{f}$ is shown in fig. 2, and 
consists of five complete branches.
The domain is accordingly partitioned into five subsets, which 
we will denote respectively by $1,2,0,3,4$. Branches $1,2,3,4$ have a 
constant absolute value of the slope $\Lambda$, while in region $0$ the 
map takes the usual Manneville and Pomeau form{\rf{MP}} 
$x_{n+1}=x_n+a\cdot x_n^{\gamma}$ ($\gamma >1$), 
leading to intermittency.
This partition induces a good symbolic dynamics:
every possible combination of symbols is physically realized as a 
trajectory of the map.

To apply the formalism we have to 
accomplish two things: enumerate all possible cycles once the marginal 
fixed point is pruned away (the effect of the marginal 
fixed point is nonlinear and is probed by the infinity of cycles which 
accumulate to it{\rf{AAC2}}), and specify $\sigma_p$ and 
$|\Lambda_p|$ for each cycle. It is easy to see that the symbolic 
dynamics corresponding to prohibiting an infinite repetition of $0$
is determined by 
unrestricted grammar in the following (countable) alphabet $\{ 
0^k1,0^l2,0^n3,0^i4,1,2,3,4\}$ $(k,l,n,i=1,2,3\dots)$. As regards 
winding numbers, if we denote by $\epsilon_j$ a generic letter of the 
alphabet we have $\sigma_{\epsilon_1 \epsilon_2 \dots 
\epsilon_n}=\sum_{i=1}^n\,\sigma_{\epsilon_i}$ where for each letter 
including
symbols $1$ or $2$ $\sigma_{\epsilon}=1$, while for each letter 
including
symbols $3$ or $4$ $\sigma_{\epsilon}=-1$. We now use the piecewise 
linear approximation of Gaspard and Wang{\rf{INT}} to the Manneville 
Pomeau system to estimate the stability of cycles shadowing the 
marginal fixed point, so that
$|\Lambda_{\epsilon_1 \epsilon_2 \dots \epsilon_n}|=\prod_{j=1}^n 
|\Lambda_{\epsilon_j}|$, and for the various letters we have
\beq \begin{array}{r@{\,=\,}l@{\;\;\;\;\;}l}
|\Lambda_{\epsilon}| & \Lambda &\epsilon=1,2,3,4 \\
|\Lambda_{\epsilon}| & \Lambda\cdot k^{\alpha+1} 
&\epsilon=0^kL\,\,L=1,2,3,4
\end{array}
\label{W}
\eeq
with $\alpha=1/(\gamma-1)$.
As each curvature{\rf{PPRL}} is zero we get that the zeta function is 
expressed in terms of the fundamental cycles{\rf{AAC1}} (which are determined by 
the letters) and takes the following form
\[
\zeta_0^{-1}(z,\beta)=1-\frac{4 \cosh \beta}{\Lambda} 
z \left(1+ \sum_{k=1}^{\infty} \frac{z^k}{k^{\alpha+1}}\right)
\]
In this way (\ref{ast}) reads
\[
\sigma^2(t)\sim t\frac{\sum_{k=1}^{t} k^{-(\alpha+1)}}{\sum_{k=1}^{t} 
k^{-\alpha}}
\]
This leads to normal diffusivity for $\alpha > 1$, while for other 
parameter values we get
\beq
\sigma^2(t) \sim
\left\{
\begin{array}{l@{\quad for \;\;}l}
t^{\alpha} & \alpha \in(0,1) \\
t/\ln t & \alpha =1
\end{array}\right.
\label{astast}
\eeq
in agreement with probabilistic estimates and numerical simulations 
in {\cite{GT}}. We remark
that our findings are entirely based on metric and 
topological properties of the system, thus the relation, expressed by 
(\ref{astast}), between anomalous diffusion and exponents of power--law 
stabilities does not rely on any
stochastic modelization of the system.

Our theory may also be applied when the diffusion is 
accelerated{\rf{GNZ}}, that is when marginal fixed points appear in 
the running branches (see fig. 3).
We will give a detailed treatment of the results elsewhere: here we 
just quote the results (in agreement with {\cite{GNZ}}, except the 
case $\alpha=1$)
\beq
\sigma^2(t)\,\sim \left\{
\begin{array}{l@{\quad for \;\;}l}
t & \alpha > 2 \\
t \ln t & \alpha =2 \\
t^{3-\alpha} & \alpha \in (1,2) \\
t^2/\ln t & \alpha =1 \\
t^2 & \alpha \in (0,1)
\end{array}\right.
\label{mm}
\eeq
where $\alpha$ is still $1/(\gamma-1)$.
We observe that topological intricacies of the system in its laminar 
phase do not affect the asymptotic estimates (\ref{astast}) and 
(\ref{mm}), which are exclusively determined by the sequence of cycles 
accumulating to the marginal fixed points. It is precisely this 
robustness with respect to fine details of the dynamics which let us 
claim that our approach may be relevant also in much more complex 
systems, like the Lorentz gas with infinite horizon. This may 
be appreciated by a 
simple argument: the weight of orbits travelling a time $t$ without 
collisions in this system goes like $t^{-3}${\rf{BUNI}}, corresponding 
to $\alpha=2$ in (\ref{W}). From (\ref{mm}) we thus get 
$\sigma^2(t)\sim t \ln t$, which indeed seems to reproduce the correct 
behaviour{\rf{UH}}.

In this paper we have proposed an extension of the periodic orbit theory of 
deterministic diffusion{\rf{DA,DCEG}}, capable of predicting 
the asymptotic time 
behaviour of $\sigma^2(t)$ by using an appropriate 
cycle expansion. In particular this technique has been tested with 
classes of one--dimensional maps in which intermittency slows or 
accelerates diffusion: the anomalous diffusion is completely 
characterized by the exponent $\alpha$ ruling power--law stability of cycles 
shadowing the marginal fixed points ($\alpha$ is in turn connected to
the
intermittency exponent $\gamma$). 
Our results provide examples in which cycle expansions can 
successfully deal with marginal stability: this supports the view that 
this technique may represent a major tool in the analysis of generic 
low--dimensional (both classical and quantum) hamiltonian systems.

{\bf Acknowledgements:} 
{We warmly thank Predrag Cvitanovi\'c and Arkady Pikovsky for useful 
comments.}
\vfill\eject
\renewcommand{\baselinestretch} {1}

\vfill\eject
{\bf {Figure captions}}
\vskip 20pt
\begin{enumerate}
\fg{f1}
Example of a circle map with a marginal fixed point.
\fg{f2}
Map on the torus associated to the one shown in fig. 1.
\fg{f3}
$\hat{f}$ for a model of accelerated diffusion: marginal fixed points 
are situated in the running branches ($|\sigma|=1$).
\end{enumerate}
\end{document}